\documentstyle[12pt]{article}

\font\twelve=cmbx10 at 15pt
\font\ten=cmbx10 at 12pt

\begin{document}

\begin{titlepage}

\begin{center}

\renewcommand{\thefootnote}{\fnsymbol{footnote}}

{\ten Centre de Physique Th\'eorique\footnote{
Unit\'e Propre de Recherche 7061} - CNRS - Luminy, Case 907}
{\ten F-13288 Marseille Cedex 9 - France }

\vspace{1 cm}

{\twelve BOUND STATES OF THE HYDROGEN ATOM IN THE PRESENCE OF A MAGNETIC
MONOPOLE FIELD AND AN AHARONOV-BOHM POTENTIAL}

\vspace{0.3 cm}

\setcounter{footnote}{0}
\renewcommand{\thefootnote}{\arabic{footnote}}

{\bf V\'{\i}ctor M. VILLALBA\footnote{permanent address: Centro de
F\'{\i}sica, Instituto Venezolano de Investigaciones Cient\'{\i}ficas IVIC,
Apdo. 21827, Caracas 1020-A Venezuela\\ e-mail
address: villalba@cpt.univ-mrs.fr, villalba@dino.conicit.ve} }

\vspace{2 cm}

{\bf Abstract}

\end{center}

In the present article we analyze the bound states of an electron in a
Coulomb field when an Aharonov-Bohm field as well as a magnetic Dirac
monopole are present. We solve, via separation of variables, the
Schr\"odinger equation in spherical coordinates and we show how the Hydrogen
energy spectrum depends on the Aharonov-Bohm and the magnetic monopole
strengths. In passing, the Klein-Gordon equation is solved.

\vspace{2,5 cm}


\bigskip

\noindent August 1994

\noindent CPT-94/P.3066

\bigskip

\noindent anonymous ftp or gopher: cpt.univ-mrs.fr

\end{titlepage}

After the appearance of the pioneering work of Dirac \cite{Dirac}, where the
notion of magnetic monopole was introduced, a large body of papers \cite
{Chandra,Wu,Goldhaber,Kazama} have been published trying to elucidate the
theoretical consequences of the existence of this hypothetical particle as
well as the experimental verification of its existence. A good theoretical
scenario for studying Dirac monopoles is to analyze the motion of electrons
in the field generated by magnetic charges. A different approach is to consider
the interaction of a monopole with a well studied quantum mechanical system
like the Coulomb field, and by analyzing the scattering of electrons or
looking
at the bound states, to be able to evaluate the possible contribution of
magnetic monopole. In this direction, the problem of a relativistic Dirac
electron in the presence of a Coulomb field, plus a magnetic monopole and the
Aharonov-Bohm potentials have been discussed by Hoang {\it et al.} \cite
{Hoang} They compute the relativistic Dirac wave function and also obtain
the bound states of the problem. Among the particularities of the energy
spectrum, Hoang {\it et al} find that there are quantum states for which the
Aharonov-Bohm contribution is absent. They solve the problem with the help
of complex coordinates, representing the group of transformations in
two-dimensional complex space, and introducing a vector-parameter notation
of the SU(2) group.

Looking at the result reported in Ref \cite{Hoang} a natural question arises:
is the strange behavior of the energy spectrum particular to
relativistic spin 1/2 particles, described by the Dirac equation, or is it
also observed when we deal with non relativistic spinless Schr\"odinger
particles, or spin zero relativistic particles.

It is the purpose of the present letter to show that there are states of
the spectrum of the non relativistic (Schr\"odinger) hydrogen atom in the
presence of a Dirac monopole and a AB potential where the Aharonov Bohm
contribution is absent. An analogous result is obtained when we work with the
Klein Gordon equation. Consequently, it is not the spin that is responsible for
the anomalous behavior on the energy spectrum. Throughout the article we equate
the speed of the light $c$ and the Planck constant $\hbar $ to unity.

The Hamiltonian of a (non relativistic) point charge $e$ of reduced mass m
in the potentials $V(r)$ and $\vec A$ can be described by means of the
Schr\"odinger equation
\begin{equation}
\label{1}\left[ \frac 1{2{\rm m}}(-i\nabla -e\vec A)^2+V(r)\right] \Psi
=H\Psi =i\partial _t\Psi
\end{equation}
where, in the present problem, $V(r)$ is the Coulomb scalar potential
\begin{equation}
\label{2}V(r)=-\frac{e^2Z}r
\end{equation}
$\vec A$ is the sum of the vector potential $\vec A_g$ associated with
the Dirac magnetic monopole \cite{Dirac}
\begin{equation}
\label{3}\vec A_g=\frac{g(1-\cos \vartheta )}{r\sin \vartheta }\hat
e_\varphi
\end{equation}
where the charge $g$ of the monopole satisfies the quantization condition
\begin{equation}
\label{4}eg=n/2
\end{equation}
and the Aharonov-Bohm potential \cite{Aharonov} $\vec A_{AB}$
\begin{equation}
\label{5}\vec A_{AB}=\frac F{2\pi r\sin \vartheta }\hat e_\varphi
\end{equation}

Since the potentials (\ref{2}), (\ref{3}) and (\ref{5}) do not depend on the
azimuthal angle $\varphi $ nor on the time, we have that the wave function $%
\Psi (\vec r,t),$ expressed in spherical coordinates, can be written as
follows
\begin{equation}
\label{6}\Psi =\Phi (r,\vartheta )e^{i(k_\varphi \varphi -Et)}
\end{equation}
where $\Phi (r,\vartheta )$ satisfies the partial differential equation
\begin{eqnarray}
\label{7}
\frac{\partial ^2\Phi }{\partial r^2}+\frac 2r\frac{\partial \Phi }
{\partial r}+\frac 1{r^2}\frac{\partial ^2\Phi }{\partial \vartheta ^2}+
\frac{\cot \vartheta }{r^2}\frac{\partial \Phi }{\partial \vartheta }-
\nonumber \\
\frac{\left(k_\varphi -[F/2\pi +g(1-\cos \vartheta )]\right) ^2\Phi }{r^2\sin
^2\vartheta }+2m(E+\frac{e^2Z}r)\Phi =0
\end{eqnarray}
eq. (\ref{7}) can be reduced to two ordinary differential equations with the
help of
\begin{equation}
\label{8}\Phi =R(r)\Theta (\vartheta )
\end{equation}
then, substituting (\ref{8}) into (\ref{7}) we arrive at
\begin{equation}
\label{9}\frac{d^2R}{dr^2}+\frac 2r\frac{dR}{dr}+2{\rm m}(E+\frac{e^2Z}r)R-%
\frac{\lambda ^2}{r^2}R=0
\end{equation}
\begin{equation}
\label{10}\frac{d^2\Theta }{d\vartheta ^2}+\cot \vartheta \frac{d\Theta }{%
d\vartheta }-\frac{\left( k_\varphi -[F/2\pi +g(1-\cos \vartheta )]\right)
^2\Theta }{\sin ^2\vartheta }+\lambda ^2\Theta =0
\end{equation}
where $\lambda ^2$ is a constant of separation. It is not difficult to see
that eq. (\ref{9}) is the same one we obtain in solving the problem of an
electron in a Coulomb field \cite{Davydov}. This is a consequence of the
form of the vector potentials $\vec A_g$ and $\vec A_{AB}$, whose
contribution appears in the equation (\ref{10}), governing the dependence of
the wave function on the angle $\vartheta .$ After introducing the variable $%
x,$
\begin{equation}
\label{11}x=\cos \vartheta
\end{equation}
we have that eq. (\ref{10}) takes the form
\begin{equation}
\label{12}(1-x^2)\frac{d^2\Theta }{dx^2}-2x\frac{d\Theta }{dx}-\left[ \frac{%
(m+qx)^2}{1-x^2}+\lambda ^2\right] \Theta =0
\end{equation}
where we have introduced the parameters $q$ and $m$,
\begin{equation}
\label{13}q=-ge,\ m=\frac{Fe}{2\pi }+ge-k_\varphi
\end{equation}
In order to solve eq. (\ref{12}) we make the ansatz
\begin{equation}
\label{14}\Theta (x)=(1-x)^{(m+q)/2}(1+x)^{(q-m)/2}W(x)
\end{equation}
then, substituting (\ref{14}) into (\ref{12}) we arrive at,
\begin{equation}
\label{15}(1-y)y\frac{d^2W}{dy^2}+\left[ m+q+1-2y(1+q)\right] \frac{dW}{dy}%
+(\lambda ^2-q)W=0
\end{equation}
where we have introduced the new variable $y,$ related to $x$ as follows,
\begin{equation}
\label{16}(1-x)/2=y
\end{equation}
the solution of eq. (\ref{15}) can be written with the help of the Gauss
hypergeometric function \cite{Magnus}
\begin{equation}
\label{17}W(y)=\ _2F_1(a,b,c,y)
\end{equation}
where, the parameters $a,$ $b$ and $c$ take the form
\begin{equation}
\label{18}a=q+\frac 12-\sqrt{q^2+\frac 14+\lambda ^2},\quad b=q+\frac 12+%
\sqrt{q^2+\frac 14+\lambda ^2},\ c=m+q+1
\end{equation}
Since the wave function $\Psi ,$ solution of the Schr\"odinger equation (\ref
{1}), should be normalizable, we have that the function $\Theta (\vartheta )$
satisfies
\begin{equation}
\label{19}\int_0^\pi \Theta (\vartheta )\Theta (\vartheta )\sin \vartheta
d\vartheta =\int_{-1}^{+1}\Theta (x)\Theta (x)dx=1
\end{equation}
The condition (\ref{19}) imposes some restrictions on the values of the
parameters given in (\ref{18}). In fact, we have that the Gauss
hypergeometric functions $_2F_1(a,b,c,y)$ reduce to polynomials when $a$ or $%
b$ is a negative integer. In effect, we have that the Jacobi polynomials are
related to the Gauss $_2F_1(a,b,c,y)$ functions as follows \cite{Magnus}
\begin{equation}
\label{20}P_n^{(\alpha ,\beta )}(x)=\frac{\Gamma (n+\alpha +1)}{n!\Gamma
(\alpha +1)}\ _2F_1(-n,n+\alpha +\beta +1,\alpha +1,\frac{1-x}2)
\end{equation}
Since the normalization condition for the Jacobi Polynomials \cite
{Gradshteyn}

\begin{eqnarray}
\label{norme}
\int^{1}_{-1} (1-x)^\alpha (1+x)^\beta P_n^{(\alpha ,\beta )}(x)
P_m^{(\alpha ,\beta
)}(x)dx= \nonumber  \\
\frac{2^{\alpha +\beta +1}}{2n+\alpha +\beta +1}\frac{\Gamma
(n+\alpha +1)\Gamma (n+\beta +1)}{\Gamma (n+1)\Gamma (n+\alpha +\beta +1)}%
\delta _{mn}
\end{eqnarray}
as well the regularity\footnote{%
This condition is related to the Hermitean character of the operator (\ref
{10}) which permits us to consider the constant of separation $\lambda^2$ as a
real quantity. The normalizability of $\Theta (x)$ could be guaranteed by
imposing the weaker conditions $\alpha >-1$, and $\beta >-1$} of the
function $\Theta (x)$ in $x=\pm 1$ require that $\alpha >0$ and $\beta >0,$
and we are interested in obtaining the most general solution of $\Theta
(\vartheta ),$ we are going to consider another solutions of eq. (\ref{15})
which are also expressed in terms of Gauss functions
$$
W_2=(1-x)^{c-a-b}\ _2F_1(c-a,c-b,c,x),\ W_3=x^{1-c}\
_2F_1(a-c+1,b-c+1,2-c,x)
$$
\begin{equation}
\label{21}W_4=x^{1-c}(1-x)^{c-a-b}\ _2F_1(1-a,1-b,2-c,x)
\end{equation}
which after imposing the condition of normalizability, reduce to Jacobi
polynomials. Therefore, the solution of eq. (\ref{12}) reads
\begin{equation}
\label{22}c_0(1-x)^{\mid m+q\mid /2}(1+x)^{\mid q-m\mid /2}P_n^{(\mid
m+q\mid ,\mid q-m\mid )}(x)
\end{equation}
where $c_0$ is a constant of normalization which can be obtained in a
straightforward way from (\ref{norme}), and the integer $n$ reads
\begin{equation}
\label{23}n=\sqrt{q^2+\frac 14+\lambda ^2}-\frac 12-\frac 12(\mid m+q\mid
+\mid q-m\mid )
\end{equation}
{}From (\ref{23}) we have that when the inequality
\begin{equation}
\label{24}\mid m\mid <\mid q\mid
\end{equation}
is satisfied, $n$ takes the form
\begin{equation}
\label{25}n=\sqrt{q^2+\frac 14+\lambda ^2}-\mid q\mid -\frac 12
\end{equation}
Analogously, we have that for
\begin{equation}
\label{26}\mid m\mid >\mid q\mid
\end{equation}
we obtain
\begin{equation}
\label{27}n=\sqrt{q^2+\frac 14+\lambda ^2}-\mid m\mid -\frac 12
\end{equation}
The energy spectrum of the hydrogen atom can be obtained from the
expression \cite{Davydov}
\begin{equation}
\label{28}E=-\frac{{\rm m}Z^2e^4}{2(N+l+1)^2}
\end{equation}
where N takes integer values
\begin{equation}
\label{29}N=0,\ 1,\ 2,\ ...
\end{equation}
and the parameter $l$ is related to $\lambda $ as follows
\begin{equation}
\label{30}l=\sqrt{\frac 14+\lambda ^2}-\frac 12
\end{equation}
Then, we have that when condition (\ref{24}) fulfills the energy spectrum
reduces to
\begin{equation}
\label{31}E=-\frac{{\rm m}Z^2e^4}{2\left( N+\sqrt{(n+1/2+\mid q\mid )^2-q^2}%
+1/2\right) ^2}
\end{equation}
Analogously, we have that when $\mid m\mid >\mid q\mid $, the energy takes
the form
\begin{equation}
\label{32}E=-\frac{{\rm m}Z^2e^4}{2\left( N+\sqrt{(n+1/2+\mid m\mid )^2-q^2}%
+1/2\right) ^2}
\end{equation}
Here some comments are in order. The presence of the Dirac monopole
introduces substantial modifications in the energy spectrum relative to
the pure Aharonov-Bohm + Coulomb case \cite{Kibler,Guha}. In fact when the
monopole is absent, the expression (\ref{24}) is never valid and eq. (\ref
{32}) reduces to
\begin{equation}
\label{33}E=-\frac{{\rm m}Z^2e^4}{2\left( N+n+\mid F/2\pi -k_\varphi \mid
+1\right) ^2}
\end{equation}
which is the energy spectrum of a particle in a Coulomb field plus
Aharonov-Bohm potential \cite{Kibler}, valid for any value of the parameter $%
m.$ A completely unexpected situation arises when the inequality $\mid m\mid
<\mid q\mid $ takes place. In this case we have
\begin{equation}
\label{34}\mid \frac{Fe}{2\pi }+ge-k_\varphi \mid <\mid ge\mid
\end{equation}
and the energy spectrum is given by eq. (\ref{31}) that does not depend on
the Aharonov-Bohm potential. This remarkable property was already observed
by Hoang {\it et al }\cite{Hoang}{\it \ }when the problem is tackled in the
framework of a relativistic spin 1/2 particle.

It is straightforward to extend the results obtained for the Schr\"odinger
case to a spin-zero relativistic particle associated with the Klein-Gordon
equation
\begin{equation}
\label{35}\left[ (-i\nabla -e\vec A)^2+{\rm m}^2\right] \Psi =H\Psi =\left[
i\partial _t-eV(r)\right] ^2\Psi
\end{equation}
which  can be solved in spherical coordinates in a similar way to the
Schr\"odinger equation. Writing the wave function $\Psi $ in the form
\begin{equation}
\label{36}\Psi (\vec r,t)=\frac 1rR(r)\Theta (\vartheta )e^{i(k_\varphi
\varphi -Et)}
\end{equation}
and substituting (\ref{36}) into (\ref{35}) we arrive at
\begin{equation}
\label{37}\left( \frac{d^2}{dr^2}-\frac{\lambda ^2-Z^2e^4}{r^2}+\frac{2Ze^2E}%
r-{\rm m}^2+E^2\right) R(r)=0
\end{equation}
and the differential equation for $\Theta (\vartheta )$ is just the same one
obtained for the Schr\"odinger case (\ref{10}). The energy spectrum can be
obtained after solving eq. (\ref{37}) and requiring the normalizability of
the wave function. The values of E are
\begin{equation}
\label{38}E=\frac{{\rm m}}{\sqrt{1+\frac{Z^2e^4}{L^2}}}
\end{equation}
where
\begin{equation}
\label{39}L=N+\frac 12+\sqrt{\lambda ^2+\frac 14-Z^2e^4},\ N=0,1,2...
\end{equation}
The above expression (\ref{39}) for $\mid m\mid <\mid q\mid $ reduces to
\begin{equation}
\label{40}L=N+\frac 12+\sqrt{(n+\mid q\mid +\frac 12)^2-q^2-Z^2e^4}
\end{equation}
Substituting (\ref{40}) into (\ref{38}), and expanding in $Ze^2$ we find
\begin{equation}
\label{41}E^{\prime }=E-{\rm m}=-\frac{{\rm m}Z^2e^4}{2\left( N+\frac 12+%
\sqrt{(n+\mid q\mid +\frac 12)^2-q^2-Z^2e^4}\right) ^2}
\end{equation}
and therefore the energy spectrum (\ref{38}) does not depend on the
Aharonov-Bohm field when $\mid m\mid <\mid q\mid .$The energy spectrum for $%
\mid m\mid >\mid q\mid $ can be obtained after replacing $\mid q\mid $ with
$\mid m\mid $ in eq. (\ref{40}) and (\ref{41}). It is not difficult to see
that eqs. (\ref{41}) and (\ref{31}) are identical. Analogous result we have
when $\mid m\mid >\mid q\mid .$This  shows that the result obtained for a
no-relativistic charged particle is also valid when we consider a
relativistic spinless Klein Gordon particle.

Regarding the energy spectra associated with the Schrodinger and
Klein-Gordon equations, it would be interesting to compare them with the one
obtained when we deal with a Dirac particle. In this case we have that the
energy spectrum reads \cite{Hoang}
\begin{equation}
\label{E}E={\rm m}\left( 1+\frac{Z^2e^4}{\left[ N+(\chi
^2-Z^2e^4)^{1/2}\right] ^2}\right) ^{-1/2}
\end{equation}
where
\begin{equation}
\chi ^2=(J+\frac 12)^2-q^2,
\end{equation}
Then, in the non relativistic limit, after making the expansion of the
energy spectrum $E$ (\ref{E}) in powers of $Z^2e$ we obtain
\begin{equation}
\label{E'}E^{\prime }=E-{\rm m}=-\frac{{\rm m}Z^2e^4}{2(N+\chi )^2}
\end{equation}
a result which slightly differs from (\ref{31}) and (\ref{32}) when the
magnetic monopole contribution is present. In fact, for $J+1/2=n+$ $\mid
m\mid +1$, we have that $E^{\prime }$ (\ref{E'}) reads
\begin{equation}
\label{esp}E^{\prime }=-\frac{{\rm m}Z^2e^4}{2(N+\sqrt{(n+\mid m\mid
+1)^2-q^2})^2}
\end{equation}
An analogous result can be obtained for $J+1/2=n+$ $\mid q\mid +1,$ after
making the substitution $q$ for $m$ in (\ref{esp}). It is noticeable that
such a discrepancy disappears when $q=0.$ In fact, in this case the energy
spectrum (\ref{esp}) reduces to the expression (\ref{33}). Also we have that
for $(J+\frac 12)^2\gg q^2$ the relativistic Dirac spectrum $E^{\prime }$
and the Schr\"odinger one take the same form. Then we have that the presence
of the Dirac monopole makes a difference. Since the Klein-Gordon and the
Schr\"odinger equations are associated with spin zero particles, we can affirm
that the coupling between the spin and the Dirac monopole strength is
responsible for the difference between the energy spectra (\ref{esp}) and (%
\ref{32}).

\medskip

\centerline{\bf Acknowledgments}

\noindent The author wishes to express his indebtedness to the Centre de
Physique
Th\'eorique for the suitable conditions of work. Also the author wishes to
acknowledge the CONICIT of Venezuela and the Vollmer Foundation for
financial support.

\end{document}